\documentclass[12pt]{iopart}

\usepackage{iopams}
\usepackage{graphicx}
\usepackage[breaklinks=true,colorlinks=true,linkcolor=blue,urlcolor=blue,citecolor=blue]{hyperref}

\begin{document}

\title[TianQin constellation stability]
{Impact of orbital orientations and radii on TianQin constellation stability}

\author{Zhuangbin Tan, Bobing Ye, Xuefeng Zhang}

\address{TianQin Reseach Center for Gravitational Physics and School of Physics and Astronomy, Sun Yat-sen University (Zhuhai Campus), Zhuhai 519082, P.R. China}

\ead{zhangxf38@sysu.edu.cn}


\begin{abstract}
TianQin is a proposed space-based gravitational-wave observatory mission to be deployed in high circular Earth orbits. The equilateral-triangle constellation, with a nearly fixed orientation, can be distorted primarily under the lunisolar perturbations. To accommodate science payload requirements, one must optimize the orbits to stabilize the configuration in terms of arm-length, relative velocity, and breathing angle variations. In this work, we present an efficient optimization method and investigate how changing the two main design factors, i.e., the orbital orientation and radius, impacts the constellation stability through single-variable studies. Thereby, one can arrive at the ranges of the orbital parameters that are comparatively more stable, which may assist future refined orbit design. 
\end{abstract}
%
%
%
%
%

\section{Introduction}

\begin{figure} \label{fig_TQ}
\begin{center}
\includegraphics[width=8cm]{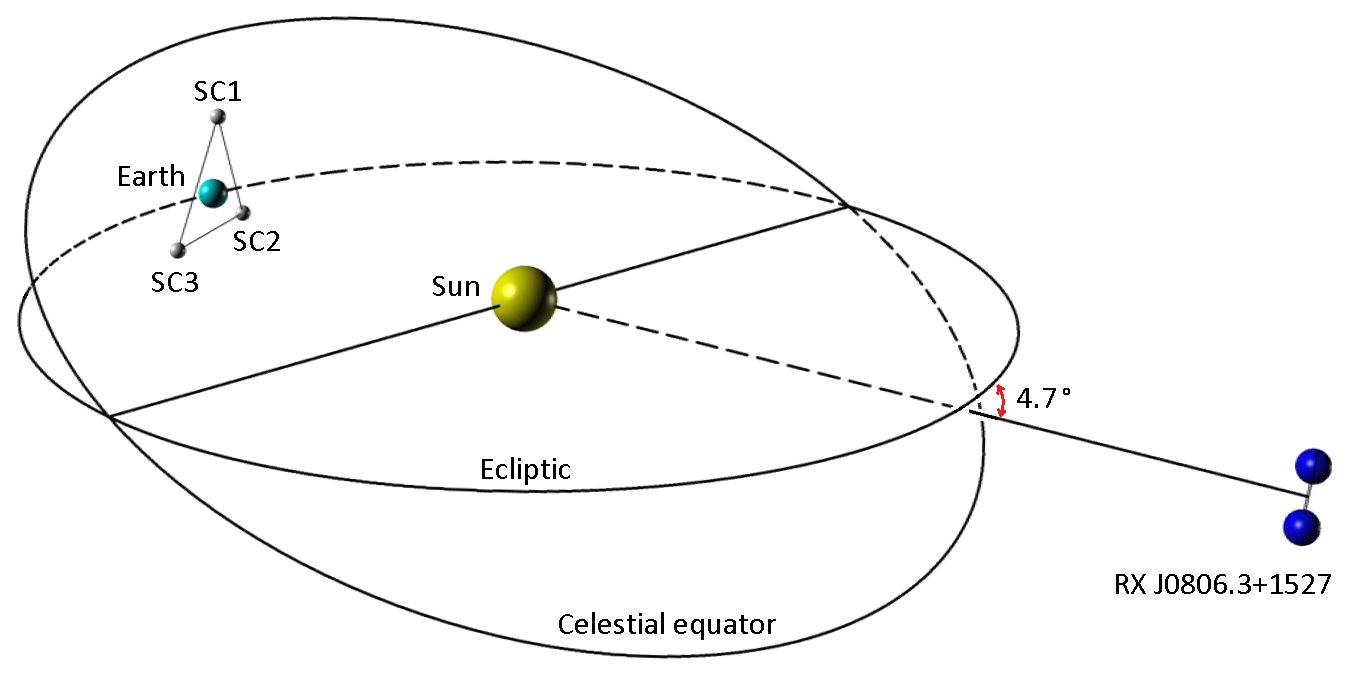}
\caption{An illustration of the TianQin constellation comprising three satellites SC1-3 (figure reproduced from \cite{Luo2016}). The direction to the reference source J0806 is shown. }
\end{center}
\end{figure}

TianQin is a geocentric space-based low-frequency gravitational-wave observatory mission consisting of three drag-free satellites in a nearly equilateral-triangle constellation \cite{Luo2016,Hu2017}. The inter-satellite measurement and inertial reference scheme is based on LISA \cite{Bender1998}. The current mission design proposes an orbital radius of $10^5$ km, and sets the orbital plane roughly perpendicular to the ecliptic plane (see Fig. \ref{fig_TQ}) and facing a verification source RX J0806.3\texttt{+}1527 (J0806 hereafter). Unlike the interplanetary LISA-like heliocentric concept \cite{Bender1998,LISA2011,NASA2012,Ni2016} with a varying constellation plane, the geocentric concept features a nearly fixed detector pointing (the normal of the constellation plane). Thus in addition to arm-lengths, there is a possibility of choosing different detector pointings, i.e., the orbital orientations, that should be accounted for in orbit design. For instance, the OMEGA mission \cite{Hiscock1997,Hellings2011} uses high circular Earth orbits as well, but near the ecliptic plane and in the retrograde direction, and has a much greater arm-length of one million kilometers. GEOGRAWI/gLISA \cite{Tinto2011,Tinto2013,Tinto2015a,Tinto2015b} and GADFLI \cite{McWilliams2011} adopt geostationary orbits in the Earth's equatorial plane with an approximately 73,000 km arm-length. The concept of LAGRANGE \cite{Conklin2011} places three spacecraft at the Earth-Moon L3, L4, and L5 Lagrange points. B-DECIGO \cite{Kawamura2018} plans to use low/medium Earth orbits (to be decided). It is worth noting that TianQin's orbit is quite different from those mentioned above (also cf. \cite{NASA2012}) in both the orientation and the radius. Hence the concept deserves careful investigations in its own right. 

Due to perturbing gravitational forces in space, the geometry of the TianQin constellation will deviate from the nominal equilateral triangle. Particularly, the long-term perturbation effects can cause persistent drift in the relative positions between the satellites, altering the overall configuration in the long run. If not restrained, the induced changes in the detector's arm-lengths and subtended angles, as well as the relative velocities between the satellites, will exceed the capacity of the science payloads (e.g., laser pointing, phase readout, laser frequency noise removal, etc.). Therefore, requirements on the constellation stability must be imposed to meet the normal working conditions of the observatory \cite{Folkner1997}. In fact, to alleviate pressure on precision instrumentation on-board, it would be desirable to have deviation from the equilateral triangle as small as possible. To summarize, acquiring stable orbits are of great necessity and importance to the mission. 

In the area of orbit stability and optimization, the heliocentric LISA-like design \cite{Vincent1987,Folkner1997,Sweetser2005} has been extensively studied with both analytic \cite{Dhurandhar2005,Nayak2006,Marchi2012,Yi2008} and numerical methods \cite{Hughes2002,Povoleri2006,Li2008,Xia2010,Wang2013a,Halloin2017,Yang2019}, over a variety of issues of perturbation analysis, arm-length and trailing-angle selection, injection error requirement, etc. Optimized orbits for the ASTROD-GW mission \cite{Ni2013} at the Sun-Earth L3, L4, and L5 Lagrange points have been obtained through tuning the average orbital periods and eccentricities \cite{Men2010,Wang2013b,Wang2015,Ni2016}. The two-week evolution of the GEOGRAWI/gLISA constellation between two station-keeping maneuvers has been analyzed in \cite{Tinto2015b}. Regarding optimization, at least two types of methods are commonly used. In the cost-function method, one minimizes a set of carefully chosen performance measures. The perturbation compensation method is to apply orbital parameter offsets to compensate for long-term perturbative effects \cite{Zhang2008,Men2010,Wan2017}. In our previous work \cite{Ye2019} for TianQin, a combined approach was developed, and seven sets of stable orbits were found with detector pointings spreading over the ecliptic plane, in addition to an earlier example presented in \cite{Luo2016} (also \cite{Hu2015}). It has been suspected that the constellation tends to be more stable with orbital planes roughly vertical to the ecliptic. Here we will put this speculation to the test. 

In this work, we intend to show how the constellation stability is affected by changing the three basic design variables, i.e., the longitude of the right ascension, the inclination, as well as the orbital radius. Thereby one may search for other orbital orientations and radii that are stable, and identify ``no-go'' parameter regions for TianQin. In addition, the work also aims to provide some insight into the perturbative effects on the constellation and the underlying mechanisms \cite{Qiao2019}. 

The paper is organized as follows. In Section 2, we describe the basic setup for the orbit simulation, including the force models and orbital parameters. In Section 3, a two-step optimization method, which includes the initial time as a variable, is presented together with the motivating ideas and comments. Section 4 exhibits our optimization results that show the impact of the three orbital elements. At the end, the concluding remarks are made in Section 5. 


\section{Simulation setup}

We simulate the satellite orbits by the open-source, flight-qualified program, General Mission Analysis Tool (GMAT) \cite{GMAT}. As in \cite{Ye2019}, the force models include a $10\times 10$ spherical-harmonic Earth gravity field (JGM-3 \cite{Tapley1996}), the point-mass gravity field from the Moon, the Sun, and seven solar system planets (the ephemeris DE421 \cite{Folkner2008}). For the drag-free mission, we assume pure gravity orbits without correction maneuvers and initial errors. Started in 2034, the propagation time is limited to a full operation cycle of one year, which we consider sufficient for the comparative studies of the orbital parameters, and also indicative of long-term (5-year) stability performance. The nominal orbits of TianQin used here are given in Table \ref{tab_nom} \cite{Luo2016, Ye2019}. The optimization variables include $a$, $i$, $\Omega$, $\nu^{\rm ini}$, as well as the initial time $t_0$, which we will discuss in the next section. 

\begin{table}[!ht]
\caption{\label{tab_nom} The nominal orbital elements of the TianQin constellation in the J2000-based Earth-centered ecliptic coordinate system (EarthMJ2000Ec). }
\begin{indented}
\item[]\begin{tabular}{@{}cccccccc}
\br
  & $a$ & $e$ & $i$ & $\Omega$ & $\omega$ & $\Delta\nu$ & $\nu^{\rm ini}$ \\ 
\mr
SC1, 2, 3 & $10^{5}$ km & 0 & $94.7^\circ$ & $210.4^\circ$ & $0^\circ$ & $120^\circ$ & $60^\circ, 180^\circ, 300^\circ$ \\
\br
\end{tabular}
\end{indented}
\end{table}


\section{Optimization method} \label{sec_method}

The combined approach used in our previous work \cite{Ye2019} consists of two mains steps. In the first step, one iteratively adjusts the initial orbital elements at a fixed initial time so that the mean semi-major axes, inclinations and longitudes of ascending nodes of the three satellites can be kept the same. In the second step, one refines the orbital elements by numerically searching for minimums of a cost function which encodes five-year stability performance. In practice, the first step runs fast, while the second suffers from long calculation time, which may take a few days depending on the numerical methods used. Hence we deem the second step inefficient for future engineering applications, as well as for the large-scale search to be performed in this work. 

\begin{figure}
\begin{center}
\includegraphics[width=8cm]{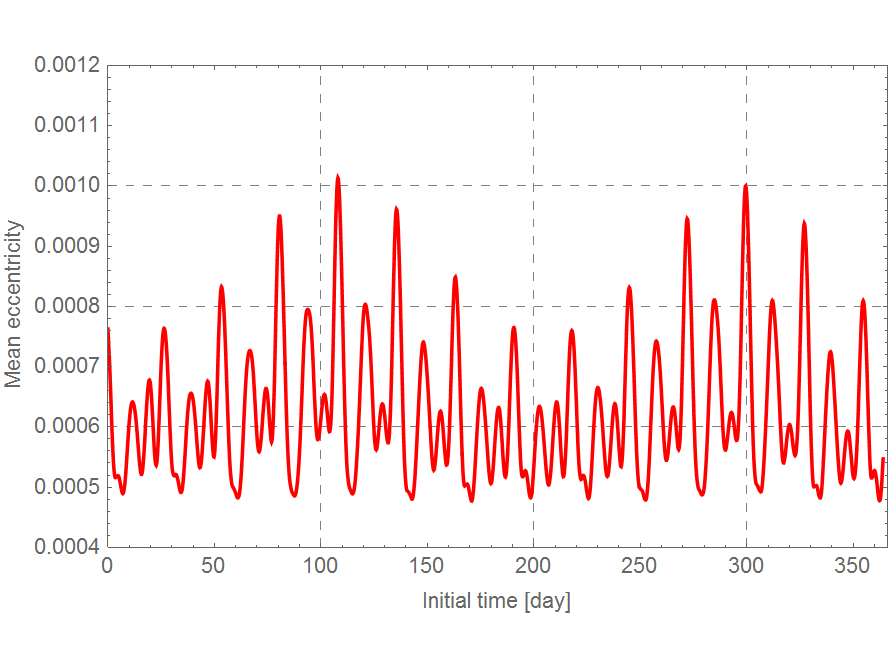}
\caption{\label{fig_t0_ecc} Mean eccentricities of one-year orbits, averaged over SC1, 2, 3 with the initial elements given in Table \ref{tab_nom}, and started at different initial times through out 2034. The curve indicates strong correlation with the lunar position such that it attains local minimums when the Earth-to-Moon vector is nearly perpendicular to the orbital plane, e.g., 22 May 2034 (day 142). }
\end{center}
\end{figure}

\begin{figure}
\begin{center}
\includegraphics[width=8cm]{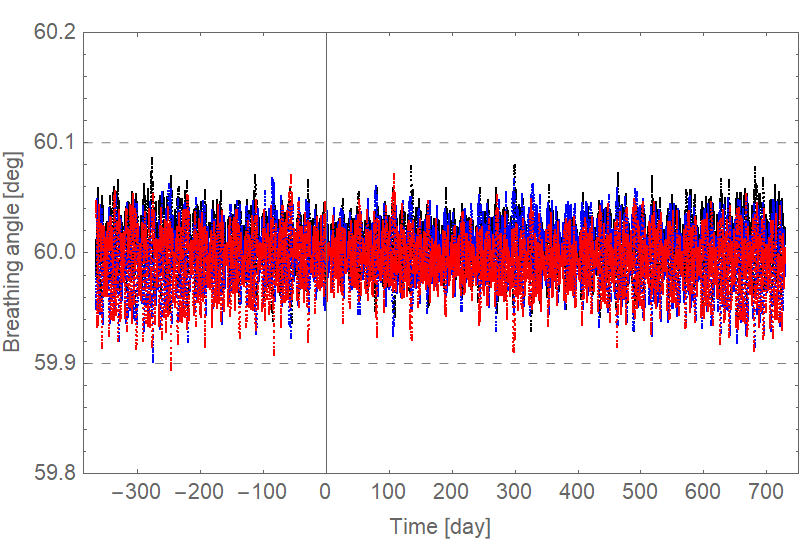}
\caption{\label{fig_backprop} Breathing angle evolutions for a set of optimized orbits (P1 of \cite{Ye2019}, Tables 4, 5 therein) backward propagated for one year from the initial time ($t=0$). The optimization is performed after $t=0$. The requirement of $\pm 0.1^\circ$ is marked by dash lines. The orbits in this case can be extended reversely for months without failing the requirement. }
\end{center}
\end{figure}

To overcome this difficulty, we propose a new method that expands on the first step of the previous approach and dispenses with the slow ``blind search'' of the cost function's optima. The idea is inspired by two observations we have made. \emph{First}, the position of the Moon (also the Sun, but less prominent), at an initial time $t=t_0$, relative to the orbital plane affects the optimization processes. It was found that for certain initial dates, such as 22 May 2034, by matching the three mean semi-major axes alone, one is able to stabilize the triangle to the required level without taking further effort. This typically happens when the Earth-to-Moon vector aligns almost perpendicularly to the orbital plane at $t_0$. To account for this phenomenon, one can look into the eccentricity evolution of the individual orbits since the stability strongly depends on the magnitude of the mean eccentricities \cite{Ye2019}. In Fig. \ref{fig_t0_ecc} we show how the mean eccentricity of one-year propagation, averaged over SC1, 2, 3 of Table \ref{tab_nom}, changes with different inputs of $t_0$. One can clearly see a monthly modulation due to the Moon. Therefore, by tuning the initial time, one can make the orbits more circular and thus obtain a more stable configuration. \emph{Second}, starting from optimized initial elements, the orbits can be backward propagated for a few months while still maintaining the same level of stability (see Fig. \ref{fig_backprop}). It allows one to extend lengths of usable orbits without the need of re-optimizing. Based on these two observations, we have decided to introduce the initial time $t_0$ into the pool of the optimization variables as the major improvement in our new method. 

The new optimization method is divided into two main steps:

\emph{Step 1}. For one-year propagation, one applies the following iteration formulas to offset the initial elements of the three satellites so that the resulting mean elements can be tuned to the desired nominal values \cite{Ye2019}. This effectively compensates for long-term linear drifts in the arm-lengths and breathing angles caused by, predominantly, the lunisolar perturbation. 
\begin{eqnarray}
a^{\rm new} = \frac{a^{\rm nom}}{\bar{a}^{\rm old}} a^{\rm old}, \\
e^{\rm new} = 0, \\
i^{\rm new} = \frac{i^{\rm nom}}{\bar{i}^{\rm old}} i^{\rm old}, \\
\Omega^{\rm new} = \Omega^{\rm old} + (\Omega^{\rm nom} - \bar{\Omega}^{\rm old}), \\
\omega^{\rm new} = 0.
\end{eqnarray}
Here for nearly circular orbits, the initial eccentricities and arguments of periapsides are set to zeros. In addition, one adjusts the true anomalies to evenly position the three satellites along the circle: 
\begin{eqnarray}
\nu_{1}^{\rm ini} = 60^\circ, \qquad \nu_{2}^{\rm ini} = 180^\circ, \qquad \nu_{3}^{\rm ini} = 300^\circ, \\
\nu_{1}^{\rm new} = \nu_{1}^{\rm old}, \\
\nu_{2}^{\rm new} = \nu_{2}^{\rm old} + [120^\circ - \bar{u}_{21}^{\rm old}], \qquad \bar{u}_{21}^{\rm old} := (\overline{\nu_{2} + \omega_{2} - \nu_{1} - \omega_{1}})^{\rm old}, \\
\nu_{3}^{\rm new} = \nu_{3}^{\rm old} + [240^\circ - \bar{u}_{31}^{\rm old}], \qquad \bar{u}_{31}^{\rm old} := (\overline{\nu_{3} + \omega_{3} - \nu_{1} - \omega_{1}})^{\rm old}.
\end{eqnarray}

\begin{figure}
\begin{center}
\includegraphics[width=8cm]{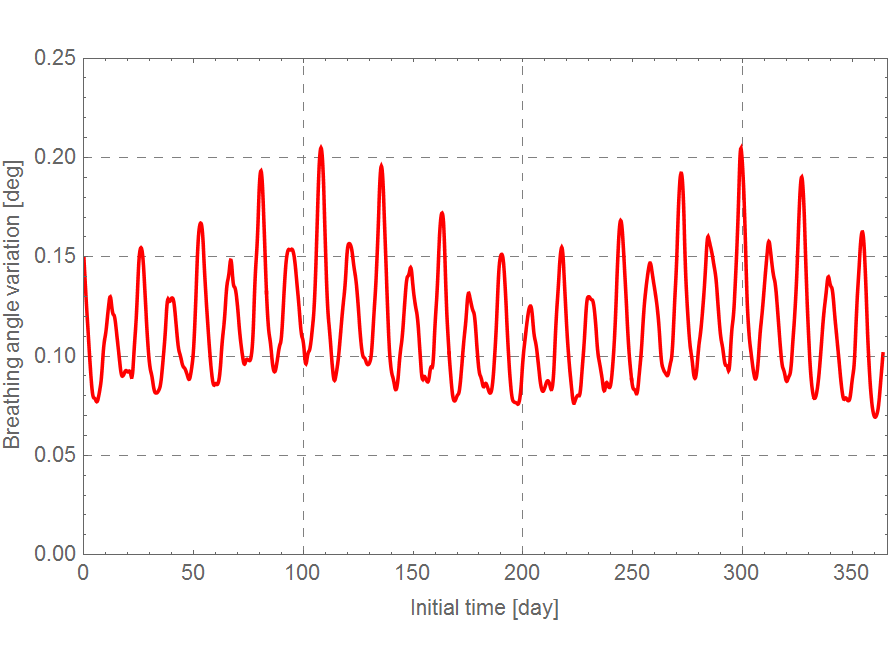}
\caption{\label{fig_t0_ang} Breathing angle variations of one-year orbits obtained from the Step 1, averaged over SC1, 2, 3 with the nominal (mean) elements given in Table \ref{tab_nom}, and started at different initial times through out 2034. The curve indicates strong correlation with the lunar position such that it attains local minimums when the Earth-to-Moon vector is nearly perpendicular to the orbital plane, e.g., 22 May 2034 (day 142). The optimized variation at the level of $\sim 0.08^\circ$ is consistent with \cite{Ye2019} using a different method. }
\end{center}
\end{figure}

\emph{Step 2}. One repeats the Step 1 for an array of initial times. For our test purposes, we have sampled $t_0$'s over the course of one year (1 Jan. 2034 - 31 Dec. 2034) and set one day apart between two adjacent times. This enables us to take into account the combined effect of initial positions of the Moon and the Sun. Then we compare all the orbit evolutions from different $t_0$'s. As discussed earlier, the essence of the Step 2 is to lower the eccentricities that lead to better stability. In Fig. \ref{fig_t0_ang}, one can see strong correlation between the breathing angle variations and the initial lunar positions (likewise for the arm-length and relative velocity variations). The variations usually attain local minimums when the initial Earth-to-Moon vector aligns upright to the orbital plane. Moreover, a half-year modulation by the Sun is also noticeable where the global minimums occur about twice a year. Thus one may select the result with the best one-year stability performance among different $t_0$'s. 

However, for our single-variable studies, we request that in all the cases the initial times be picked within the same month. This is relaxed from requesting the same initial date, which does not affect the comparisons given the second observation we made in this section (Fig. \ref{fig_backprop}). In accordance, only the local minimum (such as in Fig. \ref{fig_t0_ang}) within the requested month, not the entire year, will be used in the comparisons. 

As a comment, the entire algorithm is deterministic, and particularly, we have managed to implement the second step in parallel computing. For optimizing one set of orbital parameters, the method takes on average 2 hours to complete on 30 cores, greatly reduced from a few days of the previous approach. This is also achieved without compromising the optimizing capability, which can be seen, for instance, in the case of Fig. \ref{fig_t0_ang}. Furthermore in the Appendix, we show that the new method generates results consistent with the particle swarm optimization method (Fig. \ref{fig_PSO}).


\section{Optimization results}

Our study cases involve three orbital parameters, $i$, $\Omega$, and $a$. To make the trends clear, we only vary one parameter at a time. Without preference, the starting epoch is set in May 2034. 

\subsection{Impact of orbital orientations} \label{sec_ori}

To show the outcome of changing the orbital orientation, four separate cases are investigated with the orbital parameters given in Table \ref{tab_ori}. More specifically, we consider two orthogonal orbital planes with both prograde and retrograde orbits. They correspond to four different values of $\Omega$ with $90^\circ$ apart. For each $\Omega$, we shift the inclination from $0^\circ$ to $180^\circ$ by $\Delta i = 5^\circ$ intervals, that is, sampling along a $180^\circ$ arc with the radius $10^5$ km. 

\begin{table}[!ht]
\caption{\label{tab_ori} The (mean) orbital parameters used for studying the impact of orbital orientations. }
\begin{indented}
\item[]\begin{tabular}{@{}cccc}
\br
$a$ & $i$ & $\Delta i$ & $\Omega$ \\
\mr
$10^{5}$ km & $0-180^\circ$ & $5^\circ$ & 30$^\circ$, 120$^\circ$, 210$^\circ$, 300$^\circ$ \\
\br
\end{tabular}
\end{indented}
\end{table}

\begin{figure}[!ht]
\centering
\begin{minipage}{7.5cm}
\includegraphics[width=7.5cm]{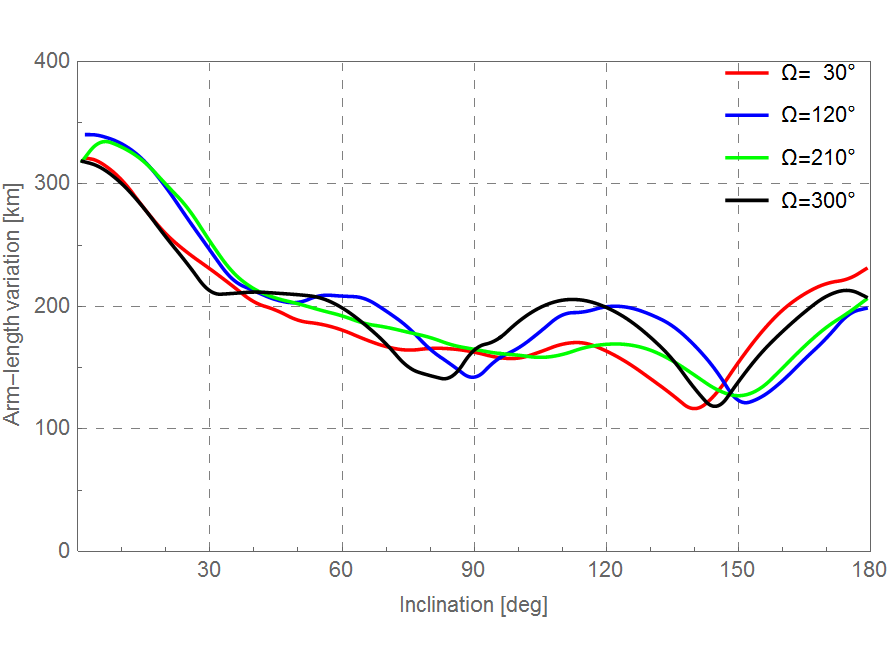}
\end{minipage}
\begin{minipage}{7.5cm}
\includegraphics[width=7.5cm]{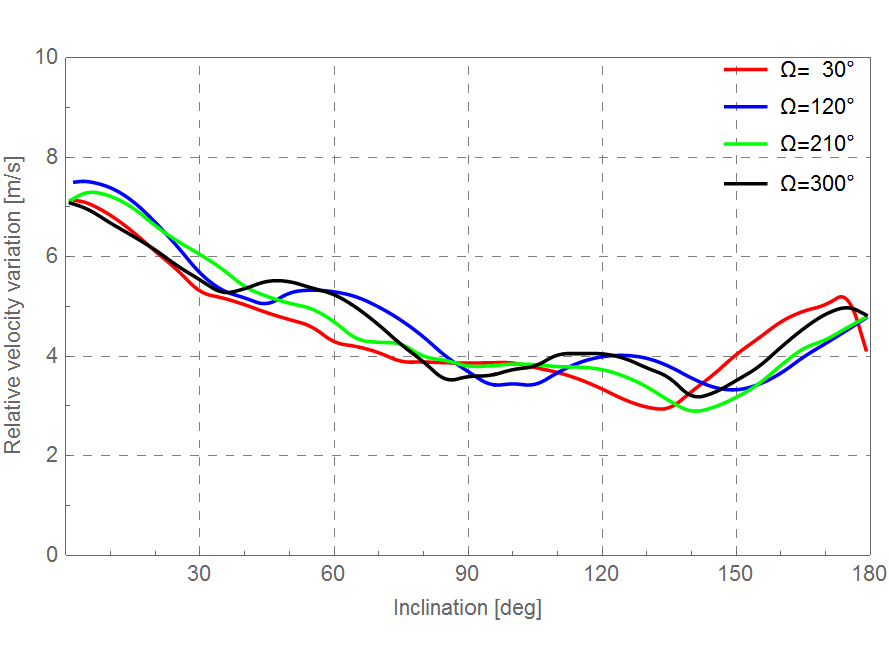}
\end{minipage}
\begin{minipage}{7.5cm}
\includegraphics[width=7.5cm]{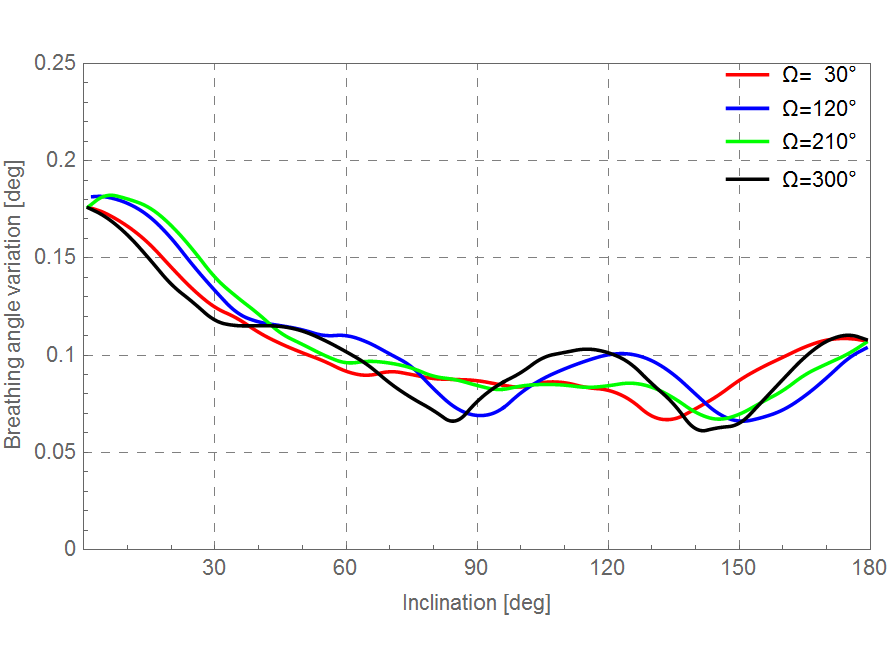}
\end{minipage}
\caption{\label{fig_orien} Impact of orbital orientations on the constellation stability, i.e., one-year variations of arm-lengths, relative velocities, and breathing angles, averaged over SC1, 2, 3, for $\Omega=30^\circ, 120^\circ, 210^\circ, 300^\circ$. The sampling interval is 5$^\circ$ in the inclination. }
\end{figure}

The results are shown in Fig. \ref{fig_orien} for the four cases. One can make a few observations here. \emph{First}, the four curves follow a similar trend. Changing $\Omega$ appears to have a relatively small impact on the constellation stability, which one may expect from an approximate rotational symmetry of the averaged perturbation about the $z$-axis in the ecliptic coordinate system. The differences in the curves, such as the trough locations and depths, can be attributed to the $5^\circ$ misalignment of the Moon's orbital plane from the ecliptic plane, and the combined initial lunisolar position relative to the orbital planes. 

\emph{Second}, the inclinations $i\approx 90^\circ$ and $i\approx 140^\circ$ stand out and exhibit favorable stability performance. It shows that, as we suspected, ``standing'' constellations are indeed comparatively more stable. 

\begin{figure}
\begin{center}
\includegraphics[width=8cm]{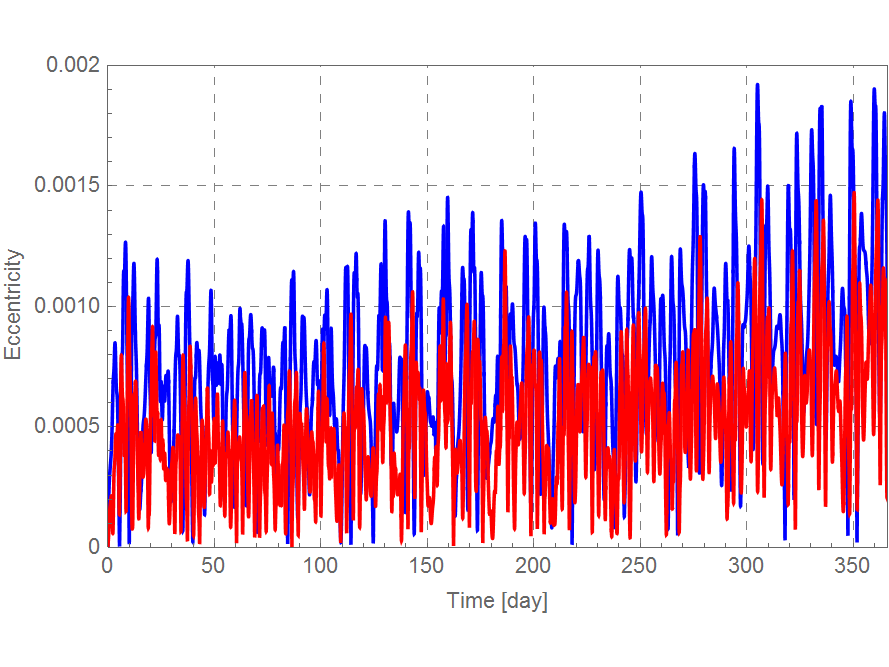}
\caption{\label{fig_ecc_pro-retro} One-year eccentricity evolutions of the retrograde orbit (red, $i=135^\circ$, $\Omega=30^\circ$) and the prograde orbit (blue, $i=45^\circ$, $\Omega=210^\circ$) lying in the same plane. The common initial elements are $a=10^5$ km, $e=0$, $\omega=0^\circ$, and $\nu=0^\circ$ at the epoch 22 May, 2034 12:00:00 UTC (EarthMJ2000Ec). The disparity of the magnitudes is evident. }
\end{center}
\end{figure}

\emph{Third}, inclined orbits moving in the retrograde direction with respect to the Moon's orbit ($i>95^\circ$) generally result in more stable constellations than those in the prograde direction ($i<85^\circ$). This also reflects from the fact that the curves in Fig. \ref{fig_orien} are lopsided and asymmetric about the mid-line $i\approx 90^\circ$. This behavior is related to irregular natural satellites, which are more commonly found in retrograde orbits. Relevant studies can be seen, e.g., in \cite{Nesvorny2003,Cuk2004}. Here, to quickly see why retrograde constellations are more stable, one can compare eccentricity fluctuations of individual orbits rotating in either direction. Indeed, as in Fig. \ref{fig_ecc_pro-retro}, we illustrate that in the same orbital plane retrograde orbits tend to have lower eccentricities than the prograde ones. 

\emph{Fourth}, orbital planes close to the ecliptic ($i\approx 0^\circ, 180^\circ$) perform unfavorably due to larger distortion per orbit by the Moon's attraction. 

\subsection{Impact of orbital radii} \label{sec_rad}

\begin{figure}[!ht]
\centering
\begin{minipage}{7.5cm}
\includegraphics[width=7.5cm]{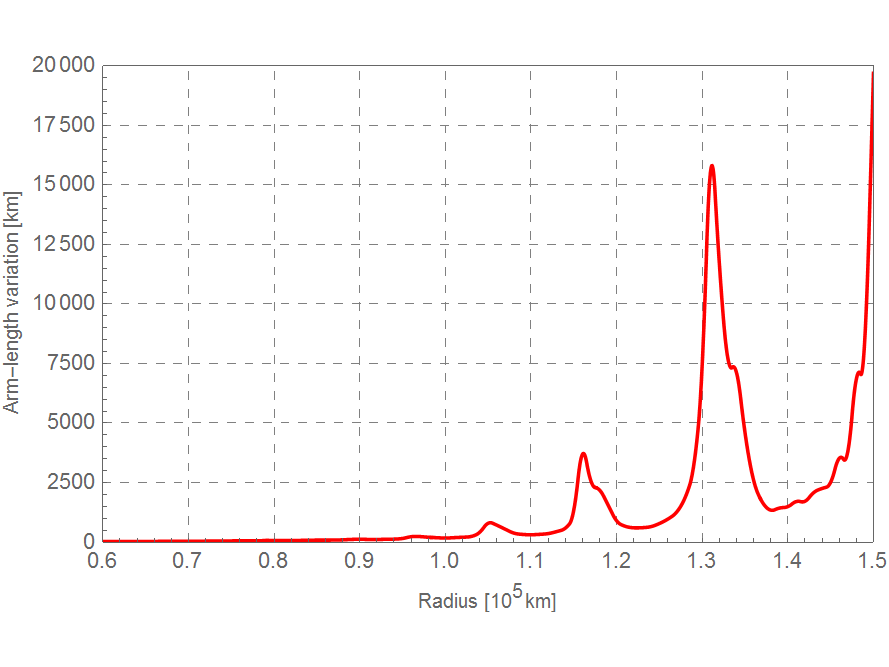}
\end{minipage}
\begin{minipage}{7.5cm}
\includegraphics[width=7.5cm]{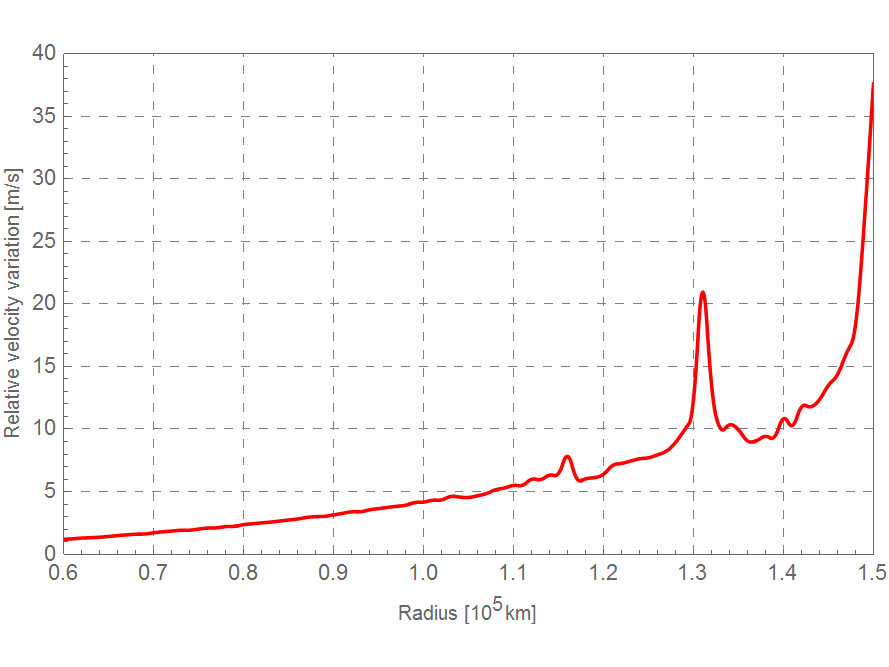}
\end{minipage}
\begin{minipage}{7.5cm}
\includegraphics[width=7.5cm]{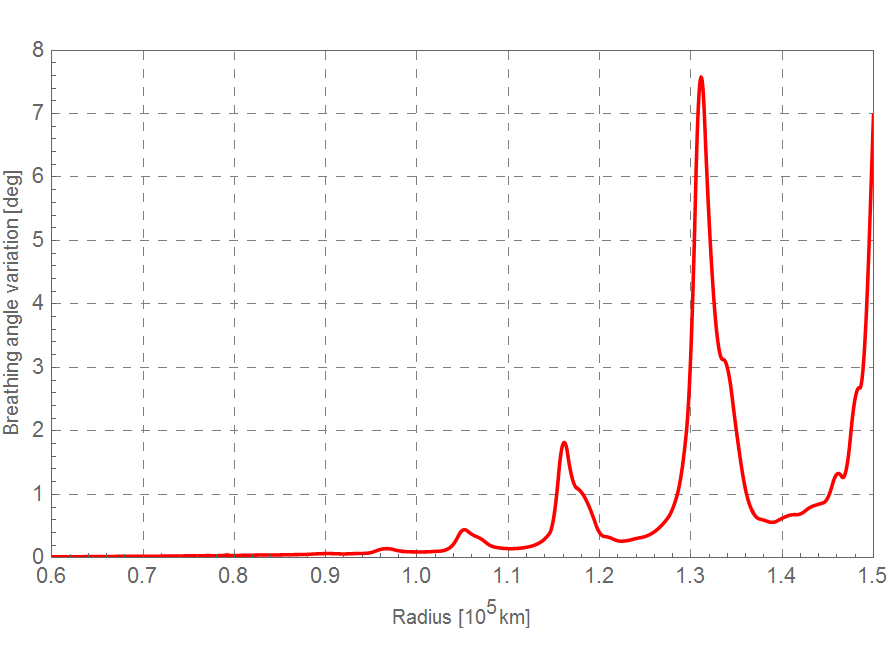}
\end{minipage}
\caption{\label{fig_radius} Impact of orbital radii on the constellation stability, i.e., variations of arm-lengths, relative velocities, and breathing angles, for the orbital plane facing J0806 ($\Omega=210.4^\circ$, $i=94.7^\circ$). The sampling interval is 1000 km. }
\end{figure}

Focused on the current TianQin orientation, i.e., the prograde orbits (w.r.t. the Earth) facing J0806 (see Table \ref{tab_nom}), we exhibit the dependence on the orbital radius from $0.6\times 10^5$ km to $1.5\times 10^5$ km in Fig. \ref{fig_radius}. 

\begin{table}[!ht]
\caption{\label{tab_ratio} The orbital period ratios between the satellites and the Moon, and the corresponding orbital radii where the resonance occurs. }
\begin{indented}
\item[]\begin{tabular}{@{}lrrrrr}
\br
$T_{\rm sat}/T_{\rm lun}$ & 1/8 & 1/7 & 1/6 & 1/5 & 1/4 \\ 
\mr
$T_{\rm sat}$ (day) & 3.4 & 3.9 & 4.6 & 5.5 & 6.8 \\
\mr
Radius ($10^4$ km) & 9.6 & 10.5 & 11.6 & 13.1 & 15.3 \\
\br
\end{tabular}
\end{indented}
\end{table}

Generally, the stability tends to worsen as the radius increases because of the Moon's attraction. On top of this trend, several peaks also show up owing to orbital resonance with the Moon. They take place at the orbital period ratio $T_{\rm sat}/T_{\rm lun}$ = $1/8$, $1/7$, $1/6$, $1/5$, $1/4$. The associated nonlinear effect severely undermines the constellation stability and cannot be mitigated to nearby non-resonant levels by the optimization method. In Table \ref{tab_ratio}, we list the radii of the resonant orbits, and one should avoid these values and their neighborhoods in orbit selection. Note that $a=10^5$ km for TianQin does not fall in the resonant regions. Our result agrees with \cite{Hu2015} by a different method. 


\section{Concluding remarks}

In this work, we have studied the influence of choosing different orbital orientations and radii on the one-year constellation stability. By using a new efficient optimization method, we can identify the ranges of the orbital parameters that show comparatively better stability. Three main conclusions can be drawn here. 

1. The constellation can persist more stably with ``standing'' (either prograde or retrograde w.r.t the Earth) orbital planes, and inclined retrograde (w.r.t the Moon) orbits with inclinations $\sim 140^\circ$, relative to the ecliptic plane. The dynamics of the latter is related to the irregular moons of the outer planets. Here we recall that OMEGA also adopts retrograde orbits \cite{Hellings2011}. In contrast, orbital planes close to the ecliptic would suffer more severely from the lunar disturbance. 

2. For a given inclination, altering the longitude of the right ascension has only a small impact on the stability. It allows a flexibility of re-orienting the detector for possible enhancement of the science output. 

3. The stability tends to degrade as the orbital radius increases. For instance, to keep the breathing angle variation within $\pm 0.2^\circ$, the orbital radius is not to exceed $1.13\times 10^5$ km. Additionally, the regions resonating with the Moon's orbit should also be avoided (Table \ref{tab_ratio}). 

The findings provide support to the initial TianQin design ($i=94.7^\circ$, $a=10^5$ km) \cite{Luo2016} and our speculation made in the introduction. From the perspective of the constellation stability, the selectable ranges of orbital orientations and radii are rather broad for TianQin, permitting further adjustment according to engineering and technological needs. In addition to the optimized orbits found in \cite{Ye2019}, other stable options, such as the inclined retrograde orbits, have been identified and may be further evaluated, from other aspects, as potential backups to TianQin. For future work, systematic refinement of TianQin geocentric orbit design will be carried out, where more environmental factors, such as solar eclipses \cite{Ye2020}, the Earth-Moon's gravity field \cite{Zhang2020}, and solar illumination, are taken into account.

\section*{Acknowledgements} 
The authors thank Dong Qiao, Jianwei Mei, Yi-Ming Hu, Jihe Wang, Defeng Gu, Yunhe Meng, Jinxiu Zhang, and Jun Luo for helpful discussion and comment. Our gratitude extends to the developers of GMAT. The work is supported by NSFC 11805287 and 11690022. 


\appendix

\section{Results compared with particle swarm optimization}

\begin{figure}
\begin{center}
\includegraphics[width=8cm]{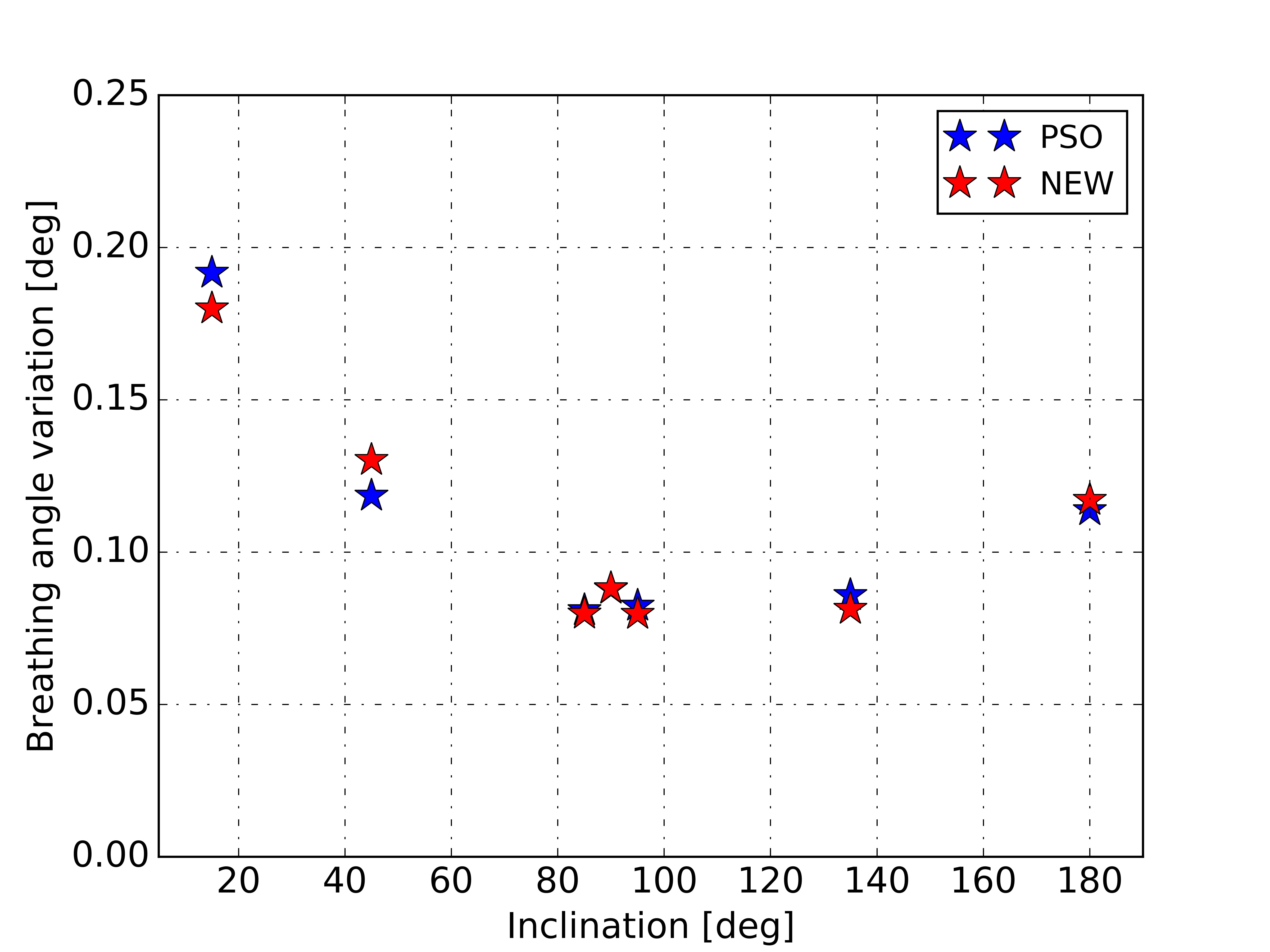}
\caption{Comparison of breathing angle variations at different inclinations obtained from PSO and the new method of Sec. \ref{sec_method}. } \label{fig_PSO}
\end{center}
\end{figure}

Particle swarm optimization (PSO) features global stochastic search of optimal solutions by a population of candidates, and has been widely used in computational science \cite{Bonyadi2017,Zhang2015}. For its implementation, we have adopted a similar cost function from \cite{Ye2019} and a swarm of 60 particles. Applied at different inclinations along the arc $\Omega = 210^\circ$, $a=10^5$ km, both methods yield consistent results and a similar trend in Fig. \ref{fig_PSO}, hence demonstrating the effectiveness of our new, and more efficient, method. 

\end{document}